\documentstyle[prd,preprint,tighten,aps,eqsecnum,%
amssymb,newlfont,epsfig]{revtex}
\begin{document}
\draft
\preprint{
\begin{tabular}{r}
UWThPh-1999-71\\
IFIC-99-84\\
November 1999
\end{tabular}
}
\title{3-Neutrino Mass Spectrum from\\
Combining Seesaw and Radiative Neutrino Mass Mechanisms}
\author{W. Grimus$^1$ and H. Neufeld$^{1,2}$}
\address{$^1$Institut f\"ur Theoretische Physik, Universit\"at Wien,\\
Boltzmanngasse 5, A--1090 Wien, Austria}
\address{$^2$Departament de F\'{\i}sica Te\`orica, IFIC, Universitat de
Val\`encia -- CSIC,\\
Apt.\ Correus 2085, E--46071 Val\`encia, Spain}
\maketitle
\begin{abstract}
We extend the Standard Model by adding a second Higgs doublet and  
a right-handed neutrino singlet with a heavy Majorana mass term. In
this model, there are one heavy and three light Majorana neutrinos with
a mass hierarchy $m_3 \gg m_2 \gg m_1$ such that that only $m_3$ is non-zero
at the tree level and light because of the seesaw mechanism, $m_2$ is
generated at the one-loop and $m_1$ at the two-loop level. We show that the
atmospheric neutrino oscillations and large mixing MSW solar neutrino 
transitions with
$\Delta m^2_\mathrm{atm} \simeq m_3^2$ and 
$\Delta m^2_\mathrm{solar} \simeq m_2^2$,
respectively, are naturally accommodated in this model without employing
any symmetry.
\end{abstract}

\section{Introduction}

At present, neutrino oscillations \cite{BP78,BP87} play a central role
in neutrino physics. Recent measurements of the atmospheric neutrino 
flux show convincing evidence for neutrino oscillations
\cite{SK-atm} with a mass-squared difference
$\Delta m^2_\mathrm{atm} \sim 10^{-3} \div 10^{-2}$ eV$^2$. It is also
likely that the solar neutrino deficit finds an explanation in terms
of neutrino oscillations \cite{sol}, either by the MSW effect \cite{MSW} with
$\Delta m^2_\mathrm{solar} \sim 10^{-5}$ eV$^2$ or by vacuum
oscillations with $\Delta m^2_\mathrm{solar} \sim 10^{-10}$ eV$^2$. For 
recent reviews about neutrino oscillations see, e.g., Ref.~\cite{review}.

Confining ourselves to 3-neutrino oscillations and thus ignoring the LSND
result \cite{LSND}, neutrino flavour mixing \cite{mixing} is described by a 
$3 \times 3$ unitary mixing matrix $U$ defined via
\begin{equation}\label{mixing}
\nu_{a L} = \sum_{j=1}^3 U_{a j} \nu_{jL}
\quad \mbox{with} \quad a = e, \mu, \tau \,,
\end{equation}
where $\nu_{a L}$ and $\nu_{jL}$ are the left-handed components of the
neutrino flavour and mass eigenfields, respectively. Then, the solar and
atmospheric neutrino mixing angles are given by
\begin{eqnarray}
\sin^2 2\theta_\mathrm{solar} & = & 
\frac{4\, |U_{e 1}|^2 |U_{e 2}|^2}%
{\left( |U_{e 1}|^2 + |U_{e 2}|^2 \right)^2}  
\label{thetasolar}\,, \\
\sin^2 2\theta_\mathrm{atm} \: & = & 4\, |U_{\mu 3}|^2 
\left( 1 - |U_{\mu 3}|^2 \right) \,,
\label{thetaatm}
\end{eqnarray}
respectively. For the small-mixing MSW solution of the solar neutrino problem,
$\sin^2 2\theta_\mathrm{solar}$ is of order $5 \times 10^ {-3}$, whereas for
the vacuum oscillation and large-mixing MSW solutions this quantity is of
order one \cite{sol}. Future experimental data will hopefully allow to
discriminate between the different possible solutions. On the other hand, for
the atmospheric neutrino oscillations the results of the Super-Kamiokande
experiment give best fit values
$\sin^2 2\theta_\mathrm{atm} = 0.99 \div 1$ and 
$\sin^2 2\theta_\mathrm{atm} \gtrsim 0.84$ at 90\% CL \cite{nakahata}.

The above-mentioned values of the
oscillation parameters pose considerable problems for
model builders in addition to the problem of
explaining the smallness of neutrino masses. From now on we
concentrate on Majorana neutrinos.
There is a vast literature on models of 3-neutrino masses and mixing
(see, e.g., the reviews \cite{altarelli,tanimoto,mohapatra} and also 
Ref.~\cite{ma} and citations therein). One possibility to explain the
smallness of the neutrino masses is the see-saw mechanism
\cite{seesaw,altarelli,lola}. The other two mechanisms are obtained by
extensions of the Standard Model (SM) in the Higgs sector \cite{konetschny}
without adding any leptonic multiplets:
The first one needs an extension by a Higgs triplet \cite{gelmini} and
leads to neutrino masses at the tree level. The smallness of the
neutrino masses is explained by the small triplet vacuum expectation
value (VEV) which is achieved by a large mass scale in the Higgs potential
(type II seesaw) \cite{scalar-seesaw}. The other possibility is given
by purely radiative neutrino masses with the generic examples of the
Zee model \cite{zee} (one-loop masses) and the Babu model \cite{babu}
(two-loop masses). Examples of these
types can be found, e.g., in Refs.~\cite{jarlskog,joshipura-2loop}.

In this paper we will discuss a model which combines the standard
see-saw mechanism with radiative neutrino mass generation. In this
framework, no other Higgs multiplets apart from scalar doublets are
needed. The most general version of such a scenario with $n_L$ lepton
doublets and charged lepton singlets, $n_R$ right-handed neutrino
singlets and $n_H$ Higgs doublets has been discussed in
Ref.~\cite{GN}. Here we confine ourselves to the most economic case
describing a viable 3-neutrino mass spectrum, namely $n_R = 1$ and 
$n_H = 2$. As was shown in Ref.~\cite{GN} (see also Ref.~\cite{tsai}), 
this case leads to a heavy
and a light neutrino at the tree level according to the see-saw
mechanism, and to one light neutrino mass at the one-loop and
the two-loop level, respectively. In the following we will demonstrate
that this model is capable of generating a hierarchical mass spectrum
fitting well with the mass-squared differences derived from the solar
MSW effect and atmospheric neutrino data and that it naturally
accommodates large mixing angles corresponding to both mass-squared
differences. The fact that tree level and loop neutrino masses appear
in our model has an analogy with the models combining the
Higgs triplet mechanism with radiative neutrino masses
\cite{ma,joshipura,typeII}. 

\section{The model}

We discuss 3-neutrino oscillations in the framework of an
extension of the SM, where a second 
Higgs doublet ($\Phi_\alpha$, $\alpha = 1,2$) 
and a right-handed neutrino singlet $\nu_R$ are present 
in addition to the SM multiplets \cite{GN}. Thus the
Yukawa interaction of leptons and scalar fields is given by
\begin{equation}\label{LY}
-\mathcal{L}_Y = \sum_{\alpha = 1}^2 
(\bar L \Gamma_\alpha \Phi_\alpha \ell_R +
 \bar L \Delta_\alpha \tilde \Phi_\alpha \nu_R) + \mbox{h.c.}
\end{equation}
with $\tilde \Phi_\alpha = i \sigma_2 \Phi^*_\alpha$. $\Gamma_\alpha$ and
$\Delta_\alpha$ are $3 \times 3$ and $3 \times 1$ matrices, respectively.
The singlet field $\nu_R$ permits the construction of an explicit Majorana
mass term
\begin{equation}\label{LM}
\mathcal{L}_M = \frac{1}{2} \,M_R\, \nu^T_R C^{-1} \nu_R + \mbox{h.c.} \,,
\end{equation}
where we assume $M_R > 0$ without loss of generality.

In this 2-Higgs doublet model, spontaneous symmetry breaking of the SM gauge
group is achieved by the VEVs
\begin{equation}\label{VEV}
\langle \Phi_\alpha \rangle_0 = \left( \begin{array}{c}
   0 \\ v_\alpha/\sqrt{2} \end{array} \right) \,,
\end{equation}
which satisfy the condition 
\begin{equation}
v \equiv \sqrt{|v_1|^2 + |v_2|^2} \simeq 246 \; \mbox{GeV} \,.
\end{equation}
The VEVs $v_{1,2}$ generate the tree level mass matrix
\begin{equation}
M_\ell = \frac{1}{\sqrt{2}} \sum_{\alpha =1}^2 v_\alpha \Gamma_\alpha
\end{equation}
for the charged leptons diagonalized by
\begin{equation}\label{Ml}
{U_L^\ell}^\dagger M_\ell U^\ell_R = \hat M_\ell
\end{equation}
with unitary matrices $U_L^\ell$, $U^\ell_R$ and with a diagonal, 
positive $\hat M_\ell$. The most general Majorana neutrino mass term 
in the model presented here has the form
\begin{equation}
\frac{1}{2} \, \omega^T_L C^{-1} M_\nu \, \omega_L + \mbox{h.c.}
\quad \mbox{with} \quad
\omega_L = \left( \begin{array}{c} \nu_L \\ (\nu_R)^c 
\end{array} \right) \,.
\end{equation}
The left-handed field vector $\omega_L$ has four entries according to the three
active neutrino fields plus the right-handed singlet. The symmetric 
Majorana mass matrix $M_\nu$ is diagonalized by
\begin{equation}
U_{\nu}^T M_\nu U_{\nu} = \mbox{diag}\, (m_1, m_2, m_3, m_4)
\end{equation}
with a unitary matrix $U_{\nu}$ and $m_i \ge 0$.

\section{The tree-level neutrino mass matrix}

From Eqs.(\ref{LY}), (\ref{LM}) and (\ref{VEV})
we obtain the tree-level version of $M_\nu$:
\begin{equation}
\label{treemass}
M_\nu^{(0)} =
\left( \begin{array}{cc}
0 & M_D^* \\ M_D^\dagger & M_R 
\end{array} \right)
\end{equation}
with
\begin{equation}
M_D = \frac{1}{\sqrt{2}} \sum_{\alpha = 1}^2 v^*_\alpha \Delta_\alpha \,.
\end{equation}
The tree-level mass matrix (\ref{treemass}) is diagonalized by the unitary
matrix
\begin{equation}\label{U0}
U_{\nu}^{(0)} = \left( u_1, u_2, u_3, u_4 \right)
\end{equation}
with
\begin{equation}\label{u}
u_{1,2} = \left( \begin{array}{c} u'_{1,2} \\ 0 
\end{array} \right), \quad 
u_3 = i\left( \begin{array}{c} \cos \vartheta\, u'_3 \\ 
\! -\sin \vartheta \end{array} \right), \quad 
u_4 = \left( \begin{array}{c} \sin \vartheta\, u'_3 \\ 
\cos \vartheta \end{array} \right),
\end{equation}
where
\begin{equation}
\tan 2\vartheta = \frac{2 m_D}{M_R}, \quad m_D = \| M_D \| =
\sqrt{M_D^{\dagger} M_D} \, .
\end{equation}
The $u'_{1,2,3}$ form an orthonormal system of complex 3-vectors with
the properties
\begin{equation}\label{u'}
u'_{1,2} \, \bot \, M_D, \quad u'_3 = M_D/m_D \,.
\end{equation}
The two non-vanishing mass eigenvalues are given by
\begin{equation}
m_3 = \sqrt{\frac{M^2_R}{4} + m^2_D} - \frac{M_R}{2} \simeq \frac{m^2_D}{M_R}
\quad \mbox{and} \quad
m_4 = \sqrt{\frac{M^2_R}{4} + m^2_D} + \frac{M_R}{2} \simeq M_R \, ,
\end{equation}
where the approximate relations refer to the limit $m_D \ll M_R$.

\section{One-loop corrections and the neutrino mass spectrum}

By one-loop corrections, the form of the neutrino mass matrix is changed to
\begin{equation}\label{M1}
M^{(1)}_\nu = \left( \begin{array}{cc}
\delta M & M_D^* \\
M^\dagger_D    & M_R \end{array} \right)
\end{equation}
with $\delta M$ being a symmetric $3 \times 3$ matrix. Its explicit form is
given by \cite{GN}
\begin{equation}\label{1loop}
\delta M = \frac{M_R}{8\pi^2} \, \sum_b \mathcal{M}^*_b \,
\frac{M^2_b \ln \left( M_R \big/ M_b \right)}{M^2_R - M^2_b} \,
\mathcal{M}^\dagger_b + M_D^* \mathcal{A} M^\dagger_D 
\end{equation}
with
\begin{equation}
\mathcal{M}_b = 
\frac{1}{\sqrt{2}} \sum_{\alpha=1}^2 b^*_\alpha \Delta_\alpha \,.
\end{equation}
In deriving this formula, terms suppressed by a factor of order $M_D/M_R$ have
been neglected. The first term in (\ref{1loop}) is generated by neutral Higgs
exchange. The sum in (\ref{1loop}) runs over all physical neutral
scalar fields 
$\Phi^0_b = \sqrt{2} \sum_{\alpha=1,2} \mbox{Re} (b_\alpha^* \Phi_\alpha^0)$,
which are
characterized by three two-dimensional complex unit vectors $b$ \cite{GN} with
\begin{equation}
\sum_b b_\alpha b_\beta = \frac{v_\alpha v_\beta}{v^2} \, .
\end{equation}
Note that we do not consider corrections to $M_D$ and $M_R$ in the neutrino
mass matrix.
The unitary matrix
diagonalizing (\ref{M1}) can be written in the form \cite{GN}
\begin{equation}\label{U1}
U_{\nu}^{(1)} = U_{\nu}^{(0)} V
\end{equation}
with $V - 1$ being of one-loop order. By an appropriate choice of the
matrix $V$ we obtain
\begin{equation}\label{M1diag}
\hat{M}^{(1)}_\nu = U_{\nu}^{(1) T} M_{\nu}^{(1)} U_{\nu}^{(1)} =
\left( \begin{array}{cccc}
{u'_1}^T \delta M u'_1 & {u'_1}^T \delta M u'_2 & 0 & 0 \\
{u'_2}^T \delta M u'_1 & {u'_2}^T \delta M u'_2 & 0 & 0 \\
0 & 0 & m_3 & 0 \\
0 & 0 & 0 & m_4
\end{array} \right).
\end{equation}
The second term in (\ref{1loop}) containing the matrix $\mathcal A$ 
(contributions from $Z$ exchange and contributions from neutral scalar
exchange other than the first term in Eq.(\ref{1loop}))
cannot contribute to (\ref{M1diag}) because
of (\ref{u'}).
The remaining off-diagonal elements in (\ref{M1diag}) can be removed by
choosing $u'_1$ orthogonal to $\Delta_1$ and $\Delta_2$. This shows at the
same time that one of the neutrinos remains still massless at the
one-loop level. However, there is no symmetry enforcing $m_1 = 0$ and the
lightest neutrino will in general get a mass at the two-loop level
\cite{2-loop}. 
Finally, the vector $u'_2$ has to be orthogonal to $u'_1$ and $u'_3$,
and its phase is fixed by the positivity of $m_2$.
Defining
\begin{equation}
c_\alpha = \frac{v}{\sqrt{2}\, m_D}\, {u'_2}^\dagger \Delta_\alpha \, ,
\end{equation}
the relation $m_D = \| M_D \|$ implies
$v_1^* c_1 + v_2^* c_2 = 0$, but the quantity 
$|c_1|^2 + |c_2|^2$ remains an independent parameter of our model,
only restricted by ``naturalness'', which requires that it is of order 1.
From (\ref{1loop}), using the Cauchy--Schwarz inequality and 
$\| b \| = 1$, we obtain the upper bound
\begin{equation}\label{m2bound}
m_2 \leq \frac{m_D^2}{8\pi^2 M_R}
\sum_b \frac{\ln \left( M_R \big/ M_b \right)}%
{1 - M_b^2 \big/ M_R^2} \, \frac{M_b^2}{v^2} 
\left( |c_1|^2 + |c_2|^2 \right) \,.
\end{equation}
Note that cancellations in Eq.(\ref{1loop}) in the summation over the
physical neutral scalars do not happen in general because the vectors
$b$ are connected with the diagonalizing matrix of the mass matrix of
the neutral scalars. The elements of these matrix are independent of the 
masses $M_b^2$ (see, e.g., Ref.~\cite{branco}).
From these considerations it follows that the order of magnitude
of $m_2$ can be estimated by
\begin{equation}\label{m2}
m_2 \sim \frac{1}{8\pi^2}\, m_3\, \frac{M_0^2}{v^2} 
\ln \frac{M_R}{M_0} \,,
\end{equation}
where $M_0$ is a generic physical neutral scalar mass. Note that for 
$M_0 \sim v$ the relation $m_2 \ll m_3$ comes solely from the numerical factor
$1/8\pi^2$ appearing in the loop integration.

\section{Discussion}

Let us first discuss the neutrino mass spectrum in the light of
atmospheric and solar neutrino oscillations. Due to the hierarchical
mass spectrum in our model we have
\begin{equation}\label{dm2}
\Delta m^2_\mathrm{atm} \simeq m_3^2 
\quad \mbox{and} \quad
\Delta m^2_\mathrm{solar} \simeq m_2^2 \,.
\end{equation}
From the atmospheric neutrino data, using the best fit value of
$\Delta m^2_\mathrm{atm}$, one gets \cite{SK-atm}
\begin{equation}\label{m3num}
m_3 = \frac{m_D^2}{M_R} \simeq 0.06 \; \mbox{eV} \,.
\end{equation}
A glance at Eq.(\ref{m2}) shows that $m_2$ is only 
one or two orders of magnitude
smaller than $m_3$ if $M_R$ represents a scale larger than the electroweak
scale. Therefore, our model cannot describe the vacuum oscillation solution of
the solar neutrino problem. On the other hand,
with the MSW solution one has \cite{sol}
\begin{equation}\label{m2num}
m_2 \sim 10^{-2.5} \; \mbox{eV}
\end{equation}
and, therefore, $m_2/m_3 \sim 0.05$, which can easily be achieved with
Eq.(\ref{m2}). In principle, the unknown mass scales $m_D$ and $M_R$ 
are fixed by Eqs.(\ref{m3num}) and
(\ref{m2num}) (see Eq.(\ref{m2}) for the analytic expression of
$m_2$). However, due to the logarithmic dependence of Eq.(\ref{m2}) on $M_R$
and the freedom of varying the scalar masses, whose natural order of
magnitude is given by the electroweak scale, the heavy Majorana mass could
be anywhere between the TeV scale and the Planck mass. 

Let us therefore give a reasonable example. Assuming that $m_D$ has something 
to do with the mass of the
tau lepton, we fix it at $m_D = 2$ GeV. Consequently, from Eq.(\ref{m3num}) 
we obtain $M_R \simeq 0.7 \times 10^{11}$ GeV. Inserting this value into
Eq.(\ref{m2}) and using (\ref{m2num}), 
the reasonable estimate $M_0 \sim 100 \div 200$ GeV
ensues, which is consistent with the magnitude of the VEVs. 
This demonstrates that our model can naturally reproduce the
mass-squared differences needed to fit the atmospheric and solar neutrino
data, where the fit for the latter is done by the MSW effect.

Now we come to the mixing matrix (\ref{mixing}), which is given by
\begin{equation}\label{mm}
U = {U_L^\ell}^\dagger U'_\nu \quad \mbox{with} \quad
U'_\nu = ( u'_1, u'_2, u'_3 )
\end{equation}
for $M_R \gg m_D$ (see Eq.(\ref{u})) and neglecting $V$ (\ref{U1}). 
Since the directions of the vectors $\Delta_{1,2}$ in the
3-dimensional complex vector space determine $U'_\nu$, we will have large
mixing angles in this unitary matrix as long as 
we do not invoke any fine-tuning
of the elements of $\Delta_{1,2}$. (This is in contrast to 
Ref.~\cite{GN} where we assumed that
$({U_L^\ell}^\dagger \Delta_\alpha)_j \sim m_{\ell j}/v$, where the
$m_{\ell j}$ are the charged lepton masses.)
Also $U_L^\ell$ might have large mixing angles, 
but could also be close to the unit matrix in analogy to the CKM matrix
in the quark sector. Since we do not expect any correlations between 
$U_L^\ell$ and $U'_\nu$, it is obvious that our model favours large
mixing angles in the neutrino mixing matrix $U$. 

On the other hand,
there is a restriction on the element $U_{e3}$ from the results of the
Super-Kamiokande atmospheric neutrino experiment and the CHOOZ result
\cite{CHOOZ} 
(absence of $\stackrel{\scriptscriptstyle (-)}{\nu}_{\hskip-3pt e}$ 
disappearance), which is approximately given by \cite{U13}
$|U_{e3}|^2 \lesssim 0.1$. Furthermore, the Super-Kamiokande results
imply that $\sin^2 2\theta_\mathrm{atm}$ is close to 1 (see introduction).
These restrictions find no explanation in our
model, but, as we want to argue, not much tuning of the
elements $U_{a3} = ({U_L^\ell}^\dagger u'_3)_a$ is needed to satisfy them. 
If we take the ratios 
$|U_{e 3}|:|U_{\mu 3}|:|U_{\tau 3}| = 1:2:2$ as an example we find
$|U_{e3}|^2 = 1/9 \simeq 0.11$ and Eq.(\ref{thetaatm}) gives
$\sin^2 2\theta_\mathrm{atm} = 80/81 \simeq 0.99$. 
To show that the favourable outcome for $\sin^2 2\theta_\mathrm{atm}$
does not depend on having $|U_{\mu 3}| \simeq |U_{\tau 3}|$, let us consider
now 1:3:2 for the elements $|U_{a3}|$. Then we obtain $|U_{e3}|^2 = 1/14
\simeq 0.07$ and $\sin^2 2\theta_\mathrm{atm} = 45/49 \simeq 0.92$.
Thus not much fine-tuning is necessary to meet the restrictions on 
$|U_{e3}|^2$ and $\sin^2 2\theta_\mathrm{atm}$ \cite{altarelli,lola,hall}. 
Obviously, our model would be in trouble if it turned out that
the atmospheric and solar neutrino oscillations decouple with
high accuracy ($U_{e3} \to 0$) or atmospheric mixing is \emph{very} close to
maximal.

Since there is no lepton number conservation in the present model, lepton
flavour changing processes are allowed. The branching ratios of the decays
$\mu^\pm \to e^\pm \gamma$ and $\mu^\pm \to e^\pm e^+e^-$ 
have the most stringent
bounds \cite{PDG}. With our assumption on the size of the Yukawa couplings 
$\Delta_\alpha$, the contribution of the charged Higgs loop \cite{BP87}
to $\mu^\pm \to e^\pm \gamma$ leads to a lower bound of about 
$100\, m_D$ for the charged Higgs mass.
The decay $\mu^\pm \to e^\pm e^+e^-$, proceeding through neutral Higgs scalars 
at the tree level,  
restricts only some of the elements of the Yukawa coupling matrices 
$\Gamma_\alpha$,
but not those of the $\Delta_\alpha$ couplings relevant in
the neutrino sector.
It is well known that the effective Majorana mass relevant in
$(\beta\beta)_{0\nu}$ decay is suppressed to a level below $10^{-2}$ eV 
in the 3-neutrino mass hierarchy \cite{betabeta}, which is considerably
smaller than the best present upper bound of 0.2 eV \cite{baudis}.

In summary, we have discussed an extension of the Standard Model with a second
Higgs doublet and a neutrino singlet with a Majorana mass being
several orders of magnitude larger than the electroweak scale. We
have shown that this model yields a hierarchical mass spectrum 
$m_3 \gg m_2 \gg m_1$ of the three light neutrinos
by combining the virtues of seesaw ($m_3$) and radiative
neutrino mass generation ($m_2 \neq 0$ and $m_1 = 0$
at the one-loop level), and that it is able to \emph{accommodate} easily 
the large mixing angle MSW solution of the solar neutrino 
problem and the $\nu_\mu\to\nu_\tau$ solution of the atmospheric 
neutrino anomaly. By construction, the neutrino sector of our model 
is very different from the charged lepton sector. The model offers 
no explanation for the mass spectrum of the charged leptons. 
We want to stress that the scalar sector of the model is exceedingly
simple and that -- apart from the Standard Model 
gauge group -- no symmetry is
involved. The moderate smallness of $|U_{e3}|^2$ and closeness of
$\sin^2 2\theta_\mathrm{atm}$ to 1 is controlled by the ratios of
the elements of the third column of the mixing matrix $U$.
We have argued that the ratios of $|U_{a3}|$ ($a=e,\mu,\tau$)
required to give 
$|U_{e3}|^2 \lesssim 0.1$ and $\sin^2 2\theta_\mathrm{atm} \gtrsim 0.84$
are quite moderate, with 1:2:2 being a good example. 
Such suitable ratios have to be assumed in the model presented here,
but might eventually find an explanation by embedding it in a larger
theory relevant at the scale $M_R$.

\end{document}